# An AI-powered Knowledge Hub for Potato Functional Genomics

**Authors:**

Yuxin Jia[1,4,*], Jinye Li[1,4], Yudong Jia[2,4], Futing Li[1], Xiaoqi Su[1], Jilin Luo[1], Yarui Dong[1], Chunyan Sun[1], Qinghan Cui[1], Li Wang[1], Axiu Li[1], Yi Shang[1], Yujuan Zhu[3,*], Sanwen Huang[3,*]

**Affiliations**

[1]Key Laboratory for Potato Biology of Yunnan Province, Joint Academy of Potato Science, Yunnan Normal University, Kunming, 650500, China.

[2]School of Information and Communication Engineering, Beijing University of Posts and Telecommunications, Beijing, 100876, China

[3]Shenzhen Branch, Guangdong Laboratory of Lingnan Modern Agriculture, Genome Analysis Laboratory of the Ministry of Agriculture and Rural Affairs, Agricultural Genomics Institute at Shenzhen, Chinese Academy of Agricultural Sciences, Shenzhen, Guangdong 518120, China.

[4]These authors contribute equally to this work.

[*]Correspondence to: jiayuxin@ynnu.edu.cn (Y.J.), zhuyujuan@caas.cn (Y.Z.), huangsanwen@caas.cn (S.H.)

**Abstract**

Potato functional genomics lags due to unsystematic gene information curation, gene identifier inconsistencies across reference genome versions, and the increasing volume of research publications. To address these limitations, we developed the Potato Knowledge Hub (http://www.potato-ai.top), leveraging Large Language Models (LLMs) and a systematically curated collection of over 3,200 high-quality potato research papers spanning over 120 years. This platform integrates two key modules: a functional gene database containing 2,571 literature-reported genes, meticulously mapped to the latest DMv8.1 reference genome with resolved nomenclature discrepancies and links to original publications; and a potato knowledge base. The knowledge base, built using a Retrieval-Augmented Generation (RAG) architecture, accurately answers research queries with literature citations, mitigating LLM "hallucination." Users can interact with the hub via a natural language AI agent, "Potato Research Assistant," for querying specialized knowledge, retrieving gene information, and extracting sequences. The continuously updated Potato Knowledge Hub aims to be a comprehensive resource, fostering advancements in potato functional genomics and supporting breeding programs.

Potato is one of the most important tuber crops. Breakthroughs in diploid potato breeding have ushered potato breeding into a new era of hybrid breeding guided by genome design. However, functional genomics research in potato still lags behind that of other major staple crops. This is primarily attributed to the lack of systematic curation and maintenance of potato functional gene information. The potato reference genome (DM1-3 516 R44) has undergone multiple iterations, with newer versions employing significantly different gene nomenclature rules compared to previous ones. This has led to widespread confusion in gene identifier (ID) citations within the literature, significantly complicating researchers' efforts to retrieve gene information. Furthermore, the number of research publications on potato has increased by approximately 100% over the past decade. This rapid growth requires researchers to undertake extensive screening and reading to keep abreast of the latest advancements, thereby increasing their workload.

The rapid development of AI technologies, particularly Large Language Models (LLMs), presents an opportunity to address these challenges. Leveraging LLMs and drawing from over 3,200 high-quality potato research papers, we have developed the Potato Knowledge Hub (http://www.potato-ai.top, Figure 1). This platform comprises two main modules: a functional gene database and a potato knowledge base. The potato functional gene database contains 2,571 non-redundant functionally genes reported in the literature, all of which have been mapped to the latest DMv8.1 reference genome[1]. Users can search using gene symbols, gene IDs from various versions, and retrieve information such as gene sequences and relevant publications. The potato knowledge base, built upon Retrieval-Augmented Generation (RAG) architecture, comprises vectorized data from these high-quality research papers[2]. It employs LLMs to answer user queries related to potato research and provide corresponding literature citations. To enhance usability for researchers, we have developed an AI agent named "Potato Research Assistant." This agent can engage in natural language conversations, discern user intent, and invoke appropriate tools to facilitate Q&A on potato knowledge, as well as gene information retrieval and sequence extraction. Furthermore, we have integrated

BLAST function into the platform, allowing users to conveniently map their genes of interest to the latest reference genome. The Potato Knowledge Hub will be continuously updated with new potato research papers and functional gene information at a frequency of at least once per month, thereby serving as a dynamic information resource for potato fundamental research and breeding.

To comprehensively acquire potato scientific research content, we searched PubMed using "potato" as a keyword in the titles and abstracts. This search, spanning from January 1, 1900, to April 30, 2025, retrieved over 32,000 papers (Figure S1). We exported the titles, abstracts, journal of publication, and DOI information for these retrieved papers and employed a LLM for filtering. The results indicated that approximately 83.5% of these papers merely mentioned the word "potato" in their titles or abstracts without being dedicated potato research articles. After this filtering process, we obtained 5,291 papers relevant to potato breeding, genetics, metabolism, physiology, genomics, and functional gene research. Recognizing that some "report" or "brief communication" type papers lack abstract information and could therefore be overlooked by LLM analysis, we manually reviewed such papers from the PubMed data export. This process allowed us to retrieve an additional 105 potato research papers.

To further refine the selection to high-quality research, we consulted the Chinese Academy of Sciences (CAS) Journal Ranking (2023 edition). We extracted potato research papers published in journals classified within the top two tiers (Q1/Q2) in the categories of Biology, Agriculture & Forestry Sciences, and Multidisciplinary Journals. Furthermore, we manually reviewed 100 recent publications on potato functional gene research to ensure comprehensive coverage. Ultimately, this process yielded a curated collection of 3,233 high-quality potato research papers. We downloaded these papers in batches, obtaining their full-text PDF files. For more accurate content retrieval and analysis, we utilized MinerU to extract text from the PDF files, converting them into plain text format. This comprehensive literature collection provides a solid foundation for constructing the Potato Knowledge Hub.

This curated collection of high-quality potato research papers provides a valuable opportunity to analyze trends in research fields and offer suggestions for future research. Using a Large Language Model (LLM), we classified this high-quality literature into 12 fields: "Genome and genetic diversity; Breeding strategies; Biotic stress response and management; Abiotic stress tolerance; Agronomic practices optimization; Tuber development, physiology and yield improvement; Nutritional quality and chemical composition; Postharvest physiology and storage technologies; Potato processing and product; Seed potato production and quality control; Sustainable agriculture; Soil health and management".

Trend analysis suggests that research related to tuber development, genome and genetic diversity, and biotic/abiotic stress responses has increased rapidly in recent decades (Figure S2). However, the proportion of papers related to tuber development has been gradually decreasing since the 1970s. This decline is likely attributable to the corresponding increase in publications on "Genome and genetic diversity" and "Abiotic stress tolerance," suggesting a shift in scientific focus towards analyzing and utilizing diverse potato germplasm (Figure S3). The proportion of research on "Biotic stress response and management" has remained stable over the past 40 years, indicating persistent efforts by researchers to address challenges posed by microbes and pests. The proportion of other research fields has shown dynamic year-to-year fluctuations, suggesting that progress in these areas has been inconsistent or challenging. Therefore, intensified future research in these dynamically changing fields could promote breakthroughs, benefiting potato breeding and the wider agricultural community.

To systematically curate potato functional gene research and address the challenge of mapping gene IDs to gene symbols across various reference genome versions, we submitted the full plain text of each paper to an LLM for analysis. The LLM was instructed to identify potato functional genes and extract their corresponding gene IDs and symbols. To ensure more comprehensive extraction, we performed two rounds of analysis. The merged results were then subjected to further manual curation, during which we reviewed 963 papers, yielding 4,221 gene ID-symbol pairs. These gene IDs originated from various databases and genome versions, including Genbank, Uniprot, SGN, ATL_v3, SolTub3.0, DMv4.03, DMv6.1, and DMv8.1[1,3,4]. We manually curated these results, extracted the corresponding amino acid sequences using the gene IDs, and mapped these IDs to the latest DMv8.1 reference genome. Finally, we harvested 2,571 non-redundant functional genes. Due to the historical lack of systematic curation of functional gene information, widespread confusion persists in gene nomenclature, exemplified by instances such as SP5G-B being incorrectly labeled as SP6A and the BZR1 gene being mistakenly named BAM7[5,6]. We manually corrected some of these results and annotated the incorrect gene symbols. To facilitate information retrieval, we have preserved the correspondence between DMv8.1 gene IDs and their gene symbols, as well as the IDs reported in the literatures. More importantly, we also retained the correspondence between gene symbols and their reporting publications. Consequently, researchers can conveniently use gene symbols or reported IDs to query detailed information about a gene in the DMv8.1 genome and access literatures that previously reported it. These efforts will greatly facilitate future potato functional gene research.

While the rapid development of Large Language Models (LLMs) offers significant convenience for scientific research, a key challenge is that most current models have not been specifically trained on potato research literature. Consequently, when answering related questions, their outputs may exhibit serious deviations due to model "hallucination." To enable researchers to query potato-related knowledge effectively, we constructed a potato knowledge base using high-quality research literature and leveraging the RAG architecture. Within this system, the LLM extracts keywords from user queries for vectorization, data retrieval, and re-ranking, subsequently answering user questions based on the retrieved and ranked information. This RAG-based approach effectively mitigates the impact of model hallucinations on the accuracy of responses, thereby assisting researchers in reliably querying specific knowledge in potato research.

To facilitate user access to the Potato Knowledge Hub, we have integrated the functional gene database and the literature knowledge base, and developed an AI agent named "Potato Research Assistant." Based on researchers' inquiries, this agent can invoke appropriate tools to summarize specific knowledge related to potato research or assist users in querying and extracting gene and promoter sequences, thereby greatly facilitating the conduct of potato scientific research.

In summary, with the assistance of Large Language Models, we have completed a systematic curation of high-quality potato research papers spanning over 120 years. Using this curated collection of high-quality potato research literature, we have developed the Potato Knowledge Hub. This platform enables researchers to use natural language to interact with an LLM for querying specialized knowledge content, extracting gene information, and retrieving relevant literature citations. The Potato Knowledge Hub is publicly accessible at www.potato-ai.top. We envision that this hub will serve as a valuable platform for functional gene research, fostering community collaboration to advance potato functional genomics and providing crucial informational support for potato breeding.

## METHODS

**Preprocessing High-quality literatures**

Potato-related literature was retrieved using the PubMed "advanced search" feature, with "Potato" as the keyword for searching titles and abstracts. Relevant studies, including their titles, abstracts, journal information, and DOIs, were exported for subsequent filtration. The title and abstract of each paper were then submitted to DeepSeek-R1 (671B) for initial screening. The system was prompted with: "Read the following literature with title and abstract, tell me if this research is focused on potato metabolism, breeding, physiology, genetics, functional genes or genomics. Answer YES or NO, do not include further explanation. Note that sweet potato (Ipomoea batatas) is not potato." Information from papers eliciting a "YES" response was extracted.

Literature lacking abstract information underwent further manual review. Papers published in Q2 or higher-ranked journals, according to the CAS Journal Ranking (2023), were retained as high-quality potato research literature. Full PDFs of these selected studies were downloaded, and their plain text was extracted using MinerU (v1.2.2).

**Potato literature knowledge base construction**

The open-source vectorization model SnowFlake (snowflake-arctic-embed-l-v2) was employed to vectorize the collected literature, with each paragraph treated as an individual chunk for this process. For data retrieval, user query keywords were vectorized using the same SnowFlake model. Cosine similarity was calculated between the vectorized keywords and each literature chunk. The top 100 chunks exhibiting the highest similarity scores were extracted. To further identify the most relevant chunks, the open-source reranking model bge-reranker-v2-m3 was utilized to assess the relevance between these retrieved chunks and the keywords. The user query, along with the top 10 most relevant chunks, was then sent to the Large Language Model (LLM) Qwen3-235B-A22B for summarization.

**Potato functional gene data base construction**

To extract gene IDs and their corresponding symbols, plain text from each piece of literature was submitted to LLMs (DeepSeek-V3-250324 and Qwen3-235B-A22B) for analysis, with instructions for the LLMs to return this information in JSON format. This extraction process involved two analytical rounds, and the resulting data were subsequently merged. All gene ID and symbol correspondences were then manually verified to correct potential inaccuracies from the LLM processing. To anchor reported gene IDs to the DMv8.1 reference genome, pairwise synteny analyses were conducted between DMv8.1 and other potato genomes (RH89-039-16, ATL_v3, Jan v1.1, SolTub3.0, DMv4.03, and DMv6.1) using MCscan within the JCVI toolkit (v1.3.6) to identify orthologs. For genes where syntenic orthologs could not be identified, and for gene sequences sourced from the Sol Genomics Network, UniProt, NCBI GenBank, and the Agria genome, orthologs in DMv8.1 were identified using the NCBI BLAST+ suite (v2.14.0) with default parameters and a sequence identity threshold of >90%. Finally, gene IDs, symbols, reported IDs, and their corresponding reference literature were compiled into an SQLite database to facilitate efficient data querying.

**Agent construction**

The agent was designed with an LLM deployed to analyze user input. If a query is determined to be unrelated to potato research, the agent politely informs the user that its scope is specialized for potato-related inquiries. For queries concerning potato research knowledge or gene information retrieval, the LLM invokes appropriate tools to fulfill the request. Python was used to write the background logical scripts, while HTML and JavaScript were employed for developing the web-based user interface.

**AUTHOR CONTRIBUTIONS**

S.H., Y.Z. and Y.J. conceived the study. Y.J. and Y.Z. wrote the manuscript. Y.J. and Y.J. developed the code for the Potato Knowledge Hub. J.L. performed the bioinformatics analyses. Y.J., J.L., F.L., X.S., J.L., Y.D., C.S., Q.C., L.W., and A.L. downloaded manuscripts and manually checked gene IDs and symbols. S.H., Y.S., Y.J., and Y.Z.

discussed and revised the manuscript. All authors read and approved the final manuscript and declare no conflicts of interest.

## ACKNOWLEDGMENTS

This work was supported by Guangdong Major Project of Basic and Applied Basic Research (2021B0301030004), the National Natural Science Foundation of China (32488302, 32302582, 32402608), Agricultural Science and Technology Innovation Program (CAAS-ZDRW202404), Shenzhen Outstanding Talents Training Fund, Yunnan Fundamental Research Projects (grant NO. 202501BC070003, 202501AT070005), China National Postdoctoral Program for Innovative Talents (BX20200376) and Yunnan Province Xingdian Talent Support Program. We extend our gratitude to all researchers who participated in the initial testing of the Potato Knowledge Hub for evaluating its functional performance and stability, and for providing their valuable suggestions.

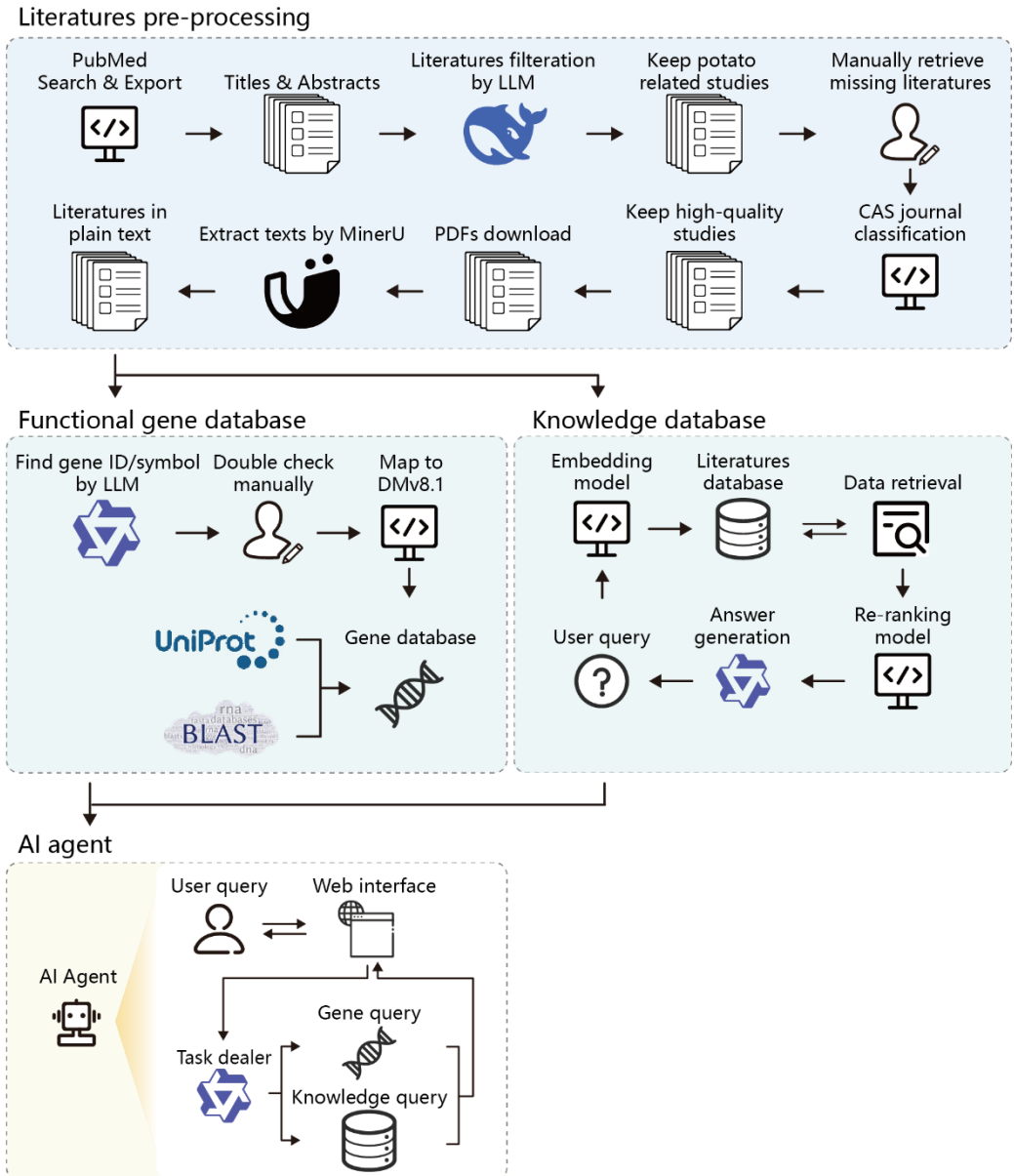

Figure 1. The work flow and architecture of Potato Knowledge Hub.

**Supplementary figures**

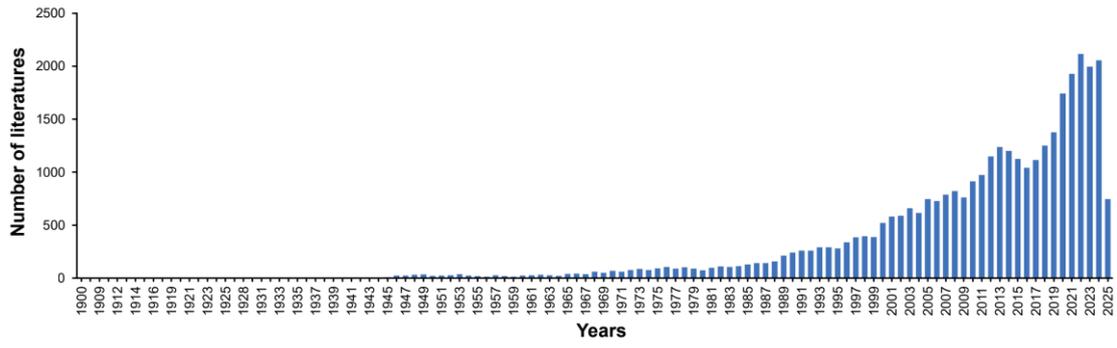

Figure S1. Number of potato related literatures published since 1900.

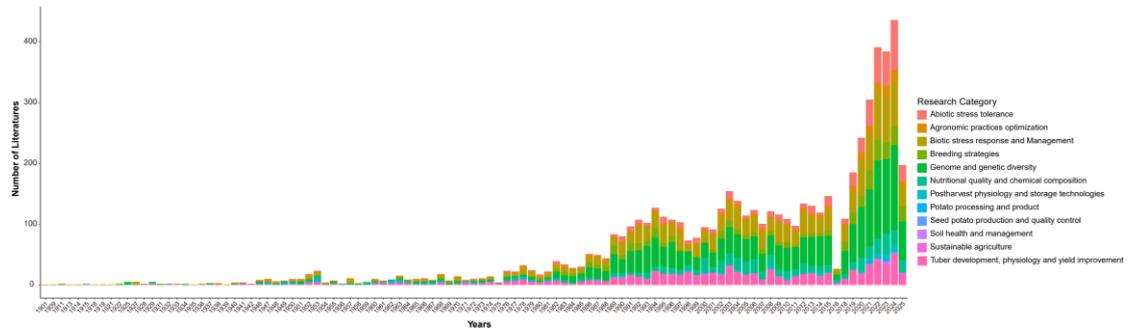

Figure S2. Trends of potato research publications across key fields.

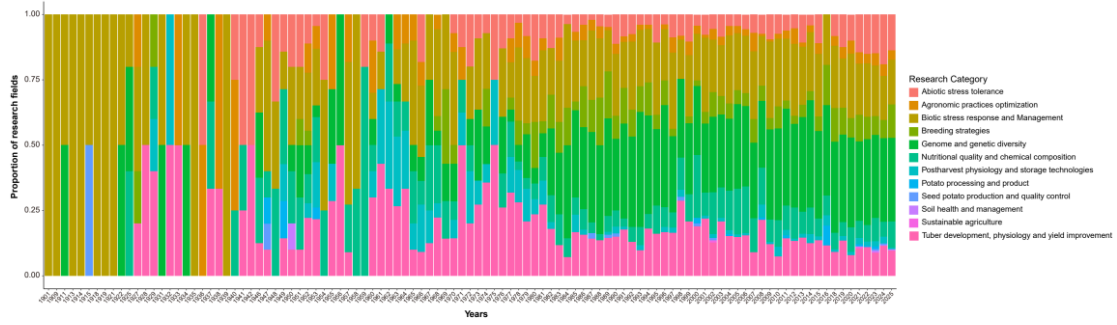

Figure S3. Trends of potato research publication proportions across key fields.